# In situ mechanical testing of an Al matrix composite to investigate compressive plasticity and failure on multiple length scales

Tianjiao Lei [a], Jenna L. Wardini [a], Olivia K. Donaldson [a], Timothy J. Rupert [a,b,*]
[a] Department of Materials Science and Engineering, University of California, Irvine, CA 92697, USA
[b] Department of Mechanical and Aerospace Engineering, University of California, Irvine, CA 92697, USA
[*] Corresponding author. Email address: trupert@uci.edu (T. J. Rupert)

SiC particle-reinforced Al matrix composites exhibit high strength, high wear resistance, and excellent high-temperature performance, but can also have low plasticity and fracture toughness, which limits their use in structural applications. This study investigates the plasticity and failure of such a composite on multiple length scales, from strain localization through a complex microstructure to the debonding of individual microparticles from the matrix. Three microscale pillars containing microstructures with different complexities and sizes/volume fraction of SiC particles were used to study the effect of these features on deformation. For the matrix, nanoscale intermetallic precipitates within the Al grains contribute most to the strengthening effect, and the Al grain boundaries are shown to be effective obstacles for preventing strain localization by dominant shear bands and, therefore, catastrophic failure. When shear localization occurs, SiC particles can then debond from the matrix if the shear band and interface are aligned. To investigate whether the interface is a weak point during catastrophic failure, a number of SiC particles were separated from the matrix with direct debonding tests, which yield an interface strength that is much higher than the critical resolved shear stress for a pillar exhibiting both shear localization and interface debonding. Therefore, the matrix-particle interface is ruled out as a possible weak point, and instead shear localization is identified as the mechanism that can drive subsequent interface debonding.

**Keywords:** Particle-reinforced composites, plasticity, failure, in situ small scale testing, matrix-particle interface

# 1. Introduction

Although lightweight metals are attractive for aerospace and automotive industries, pure Al is not used as a structural material mainly due to its low strength. Alloying must almost always be used to increase the strength of Al. In addition, ceramic reinforcements in the form of particles, fibers, or whiskers are often added to make reinforced Al matrix composites. These materials can exhibit high strength and stiffness, excellent wear resistance, and high-temperature performance, in addition to being lightweight [1,2,3]. Of the various options for reinforcement types, particles are often preferred because the microstructure and mechanical properties of composites can be controlled more precisely by varying the size and volume fraction of the reinforcement [4]. Moreover, composites reinforced by approximately equiaxed particles exhibit nearly isotropic properties, and can be formed and processed by traditional metal-working techniques with low cost [5]. SiC particles are one of the most widely used ceramic reinforcements due to their superior high-temperature mechanical properties, good thermal stability [6], excellent wear resistance, and low cost [7].

Unfortunately, the addition of reinforcements can also result in decreasing macroscopic plasticity and fracture toughness [8]. For example, by performing tensile tests on a $Al_2O_3$ particle reinforced Al matrix composite, Park et al. [9] reported that the elongation decreases from ~22% with no particles to ~9% with 30 vol.% particles. Ma et al. [10] studied the fracture mechanism for a SiC whisker reinforced Al 6061 composite by employing tensile tests, and speculated that the low plasticity of the composite may be due to a fracture process consisting of initiation and propagation of microcracks, which are dependent on the angle of loading. When the off-axis angles are small, microcracks form preferentially at the whisker ends and propagate by bypassing or pulling out the whiskers, while microcracks initiate at matrix-whisker interfaces and propagate by debonding at the interface for large off-axis angles.

The mechanical behavior of Al matrix composites is sensitive to multiple factors, such as matrix microstructure, volume fraction of reinforcements, reinforcement size and shape, and matrix-reinforcement interface bonding [11,12,13,14]. The first three factors can be studied well using a combination of advanced microscopy and macroscopic mechanical tests. However, to specifically study the local deformation near the matrix-reinforcement interface, small-scale mechanical testing techniques are necessary, with microcompression being a particular good example. Microcompression tests were developed for a better understanding of plastic deformation at small scales via samples of similar sizes with the length scale of different phases and/or dislocation activities [15,16]. Sample sizes from a few hundred nanometers to tens of micrometers in diameter are usually prepared by focused ion beam (FIB) milling. A nanoindentation system can then be employed as the mechanical test frame with a flat-punch tip to apply the load [17]. Microcompression experiments are sometimes performed in a scanning electron microscope (SEM) so that microstructural evolution can be accurately correlated with the deformation process [18], and new deformation mechanisms can be observed. For example, Greer et al. [19] reported a significant higher flow stress for single-crystalline gold micropillars than for bulk gold, and attributed the high strength to dislocation starvation. Microcompression tests are also attractive for studying the mechanical properties of nuclear materials, such as those where radiation damage is simulated with ion bombardment since in this case damage is restricted to a layer near the surface [20]. In addition, these types of tests allow for local deformation to be investigated since one can isolate interesting features and suppress fast fracture to reveal deformation details in macroscopically brittle materials [21]. For example, Khalajhedayati and Rupert [22,23] employed microcompression to nanocrystalline Ni-W to study catastrophic failure due to strain localization within shear bands, finding a correlation between strain localization and grain coarsening within the shear band. Vo et al. [24] performed microscale tensile tests on Ni-based alloys to study constrained plastic flow due to an interface and proposed a new parameter, blocked volume ratio, which is closely related to the plastic flow behavior.

In the present study, plastic deformation and failure in an Al matrix composite reinforced with SiC particles is investigated through small-scale mechanical testing, with a particular emphasis on understanding the role of the matrix-reinforcement interface. For microcompression tests, pillars of different sizes containing different microstructures were tested to elucidate the effects of different features on deformation behavior and matrix-reinforcement fracture. For direct debonding tests, six SiC particles were separated from the Al matrix to estimate the interface strength. The results show that (1) grain boundaries within the Al matrix can serve as effective obstacles for shear localization, (2) the primary strengthening of the Al matrix comes from nanoscale intermetallic precipitates, and (3) the Al alloy matrix is intrinsically strong. Most importantly, the interface strength obtained from direct debonding tests is higher than the critical resolved shear stress for a pillar exhibiting both catastrophic failure through shear localization and interface debonding. Therefore, we can conclude that the interface is not a weak point in the microstructure and instead strain localization is identified as the event that can lead to premature failure.

## 2. Materials and methods
### 2.1. Fabrication of bulk SiC particle-reinforced Al matrix composite

Fabrication of the bulk SiC particle-reinforced Al matrix composite sample was performed by standard metallurgical techniques. First, ingots of Al, Zn, Mg, and an Al-matrix intermediate alloy were used to create alloy powders using a ZW-40 atomization furnace, with powders separated by a JXZS Ultrasonic sieving machine. The composition of the alloy is Al-6Zn-1.5Mg-0.4Mn-0.2Cu-0.2Zr (wt.%). The particle diameter of the alloy powders used in subsequent steps was smaller than 75 μm. Next, the alloy powders were mixed with SiC powders in a V-shape mixing machine for 3 h. Subsequently, the alloy-ceramic powder mixture was canned, compacted under vibration, and degassed at 450 °C and $5\times10^{-3}$ Pa. Finally, the composite bulk was horizontally extruded at 450 °C. The extrusion processing was performed at a deformation rate of 0.2 mm/s and a ratio of 20:1 to yield a bulk piece with a final diameter of ~20 mm.

## 2.2. Microstructure characterization

Small samples were cut from the extruded bulk piece using a diamond saw, and then sample surfaces were mechanically polished on all sides to a fine finish. X-ray diffraction (XRD) was performed to identify the phases present using a Rigaku Smart Lab Diffractometer operated at 40 kV and 44 mA with a Cu cathode. SEM imaging and energy-dispersive spectroscopy (EDS) measurements were employed to examine the distribution of the SiC particles as well as the chemical composition of the composite using an FEI Magellan 400 XHR SEM. Conventional and scanning transmission electron microscopy ((S)TEM) paired with EDS were used to examine the structure and elemental distribution of nanoscale features using a JEOL 2800 and a JEM-ARM300F Grand Arm TEM, which were operated at 200 kV and 300 kV, respectively. Both as-prepared alloy samples and cross-sectional TEM specimens of pillars after microcompression tests were fabricated using the FIB lift-out method in an FEI Quanta 3D FEG dual beam SEM/FIB microscope equipped with an Omniprobe manipulator.

**2.3. Small-scale mechanical testing**

In situ microcompression tests were performed inside an SEM on micron-size pillars using a FemtoTools FT-NMT03 nanomechanical testing system. The load was applied by a small platen, which was FIB milled from a 50×50 µm flat Si MEMS-based force sensor-head (model FT-S200'000 with a ±200,000 µN force range and 0.5 µN force resolution). The pillar displacement was controlled using a subnanometer-resolution piezo-based actuation system. Pillars with a height-to-diameter ratio of 2-3 were fabricated to prevent plastic buckling [25] in the FEI Quanta 3D FEG dual beam SEM/FIB with a $Ga^+$ beam. Concentric annular milling was used to minimize the initial tapering effect, followed by lathe milling [26] to smooth the pillar surface and remove taper. Best practices for microcompression experiments suggest that pillars should have a near-zero taper angle to ensure a uniform stress-state during loading [25]. A thin protective Pt cap deposited on the pillar surface and decreasing ion beam current during the annular milling were employed to minimize $Ga^+$ implantation, so that $Ga^+$ effects do not play a dominant role in the plastic behavior and failure mode [27,28]. Figures 1(a)-(c) show the three pillars investigated in the present study. Pillar A contains several Al grains, one SiC particle at the top and another inside the center of the pillar, and has a variety of multiscale intermetallic precipitates inside the matrix grains. This sample can be thought of as a representative volume element of the composite, as it contains all of the important structural features found in the bulk material. Both Pillar B and C contain a SiC particle at the top and one Al grain at the bottom, with the volume fraction of the SiC particle being higher in Pillar C than in Pillar B. The SiC particle in Pillar B is located on the back side of the pillar as it is viewed in Figure 1(b). Therefore, the structure of Pillar C is the closest to isolating the interface between the Al matrix and the SiC reinforcement particle (termed the "matrix-particle interface" throughout this paper), while that of Pillar B is an intermediate state between the other two pillars. These three pillars offer a chance to investigate the deformation behavior of various microstructural complexities. The microcompression tests were displacement-controlled and performed at a nominal strain rate of $10^{-3}$ s$^{-1}$ at room temperature. The yield strength for each pillar was obtained using a common 0.2% offset criterion since Brandstetter et al. [29] verified that this criterion is still valid for ultrafine-grained Ni, and the grain size of the present composite is in micrometer range.

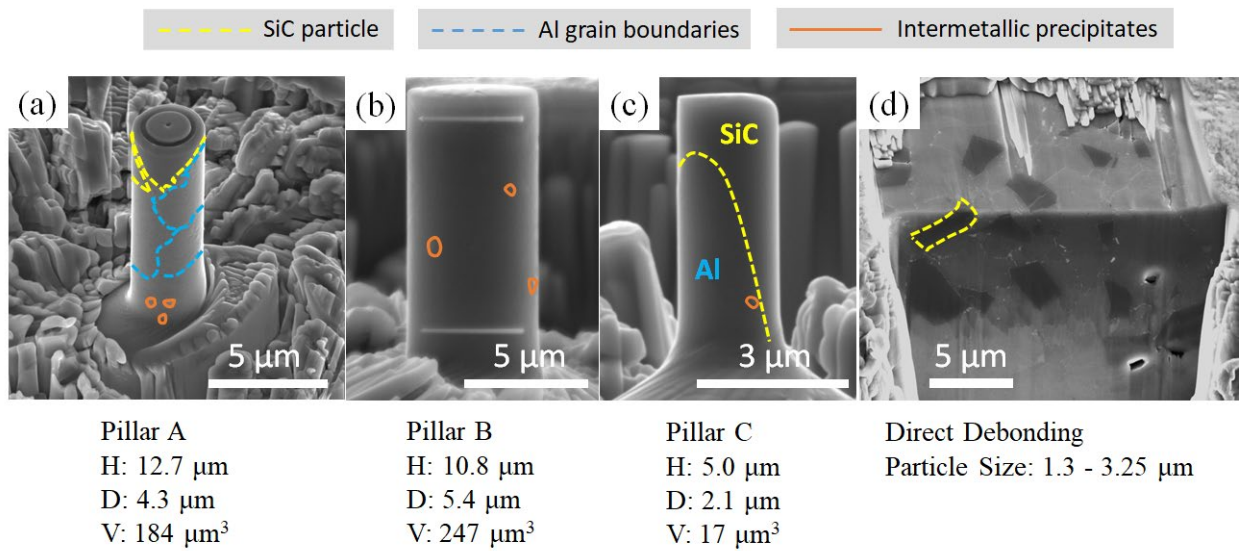

**Fig. 1 (a)-(c) SEM micrographs of three pillars with different microstructures and (d) two perpendicular surfaces milled to reveal a SiC particle. H, D, and V correspond to height, diameter, and volume, respectively. Various features including SiC particles, Al grain boundaries, and intermetallic precipitates are outlined**

Direct interface debonding experiments were also performed on individual SiC particles using the FemtoTools nanomechanical testing system to isolate the matrix-particle interface failure.

Instead of a large flat contact surface as in the microcompression tests, the force sensor head was milled to a smaller probe with a slightly rounded tip and was used to apply load to a SiC particle until the particle separated from the Al matrix. For sample preparation, two perpendicular surfaces of the composite were milled using FIB until a SiC particle was shown on both surfaces, with an example shown in Figure 1(d). The particle dimensions were determined by FIB sectioning at various depths so that the interfacial area between the matrix and particle can be determined. This technique allows for a direct measurement of interface debonding, with the interface strength determined as an average value from measurements on six particles.

## 3. Results and Discussion

Figure 2(a) shows a backscattered electron SEM micrograph of the composite surface, where SiC particles appear dark due to their relatively low effective atomic number and intermetallics comprised of higher effective atomic number constituents can be seen as bright spots on the Al-matrix background. The particles have irregular, faceted shapes with sizes ranging from sub-micrometer to a few micrometers. They tend to form clusters instead of being uniformly distributed, similar to other cast SiC particle reinforced Al matrix composites [30]. Quantitative EDS analysis of the composite microstructure shows a SiC particle volume fraction of ~10%. A magnified SEM image and the corresponding EDS maps are shown in Figure 2(b), where four particles are observed and verified to be SiC. The matrix is mainly composed of Al, along with several other alloying elements such as Zn, Mg, Cu, Mn, and Zr. Zn, Mg, and Cu are commonly used as alloying elements in Al alloys because the precipitates they form are effective for precipitation-hardening, while Mn and Zr are minor addition elements which help to nucleate precipitates [31]. In addition, small amounts of Mg and Zr have been used to improve the wettability of SiC particles in an Al-based metal matrix composite [32]. Therefore, the chemical elements of the present Al matrix are consistent with those of other conventional Al alloys and Al-based metal matrix composites.

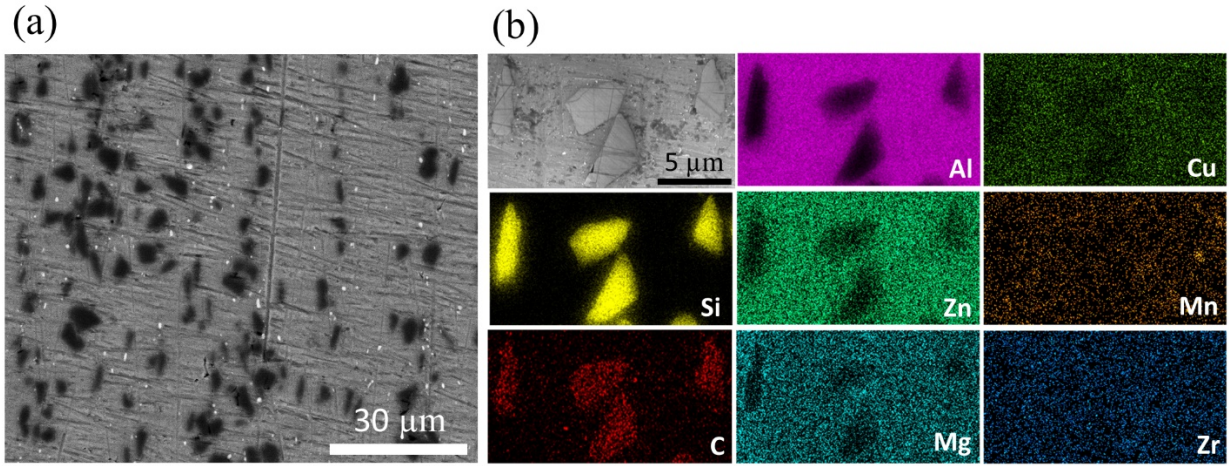

**Fig. 2 (a) Backscattered electron SEM micrograph of the composite surface, which shows SiC particles (dark regions) and intermetallics (light regions) distributed in the Al matrix. (b) A magnified SEM image and the corresponding EDS maps of different elements within the composite**

Figure 3(a) displays a representative XRD profile of the as-extruded composite, where peaks corresponding to the face-centered-cubic (FCC) Al matrix and SiC particles are clearly identified. In addition, peaks associated with various Al- and Mg-based intermetallic phases emerge, such as $MgZn_2$, which is one of the most commonly observed phases in the as-cast microstructures for commercial Al alloys [33]. To visualize the different phases, bright field TEM was employed and one example is shown in Figure 3(b). In this image, one SiC particle and an Al grain boundary are outlined in yellow and blue dashed lines, respectively, while sub-micrometer intermetallic precipitates (indicated by orange arrows) reside in the Al grain. Since the matrix microstructure is an important factor affecting the mechanical behavior of the composite, a more detailed and quantified characterization of the Al matrix was performed with bright field TEM.

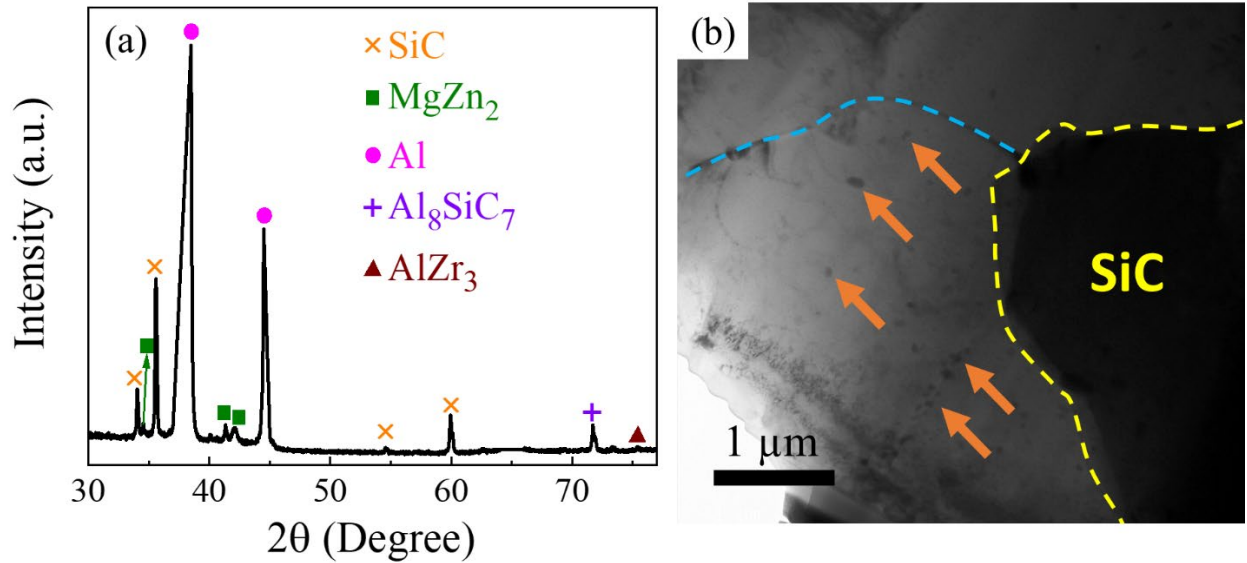

**Fig. 3 (a) XRD profile showing different phases, and (b) bright field TEM micrograph of the composite. One Al grain boundary and SiC particle are outlined by blue and yellow dashed lines, respectively. Orange arrows point to intermetallic precipitates**

Figure 4 illustrates various features and their quantification within the matrix. In Figure 4(a), an FCC Al grain is outlined whose size is ~4 μm. The inset shows a cumulative distribution function of Al matrix grain size, where the smallest and largest grain sizes are ~0.5 μm and ~4.6 μm, respectively, with an average of 2.2 μm. Therefore, the sizes of the Al grains and SiC particles (Figure 2) are similar, with the SiC particles being slightly larger. Figure 4(b) shows intermetallic precipitates within the Al grains, and a wide size range is observed from tens to hundreds of nanometers. A magnified region containing only the nanoscale precipitates is shown Figure 4(c), with quantitative characterization presented in the inset. The nanoscale precipitate sizes ranged from ~ 10 nm to 38 nm, with an average size of 23 nm. The nanoscale intermetallic precipitates were therefore much smaller than the Al grains and SiC particles.

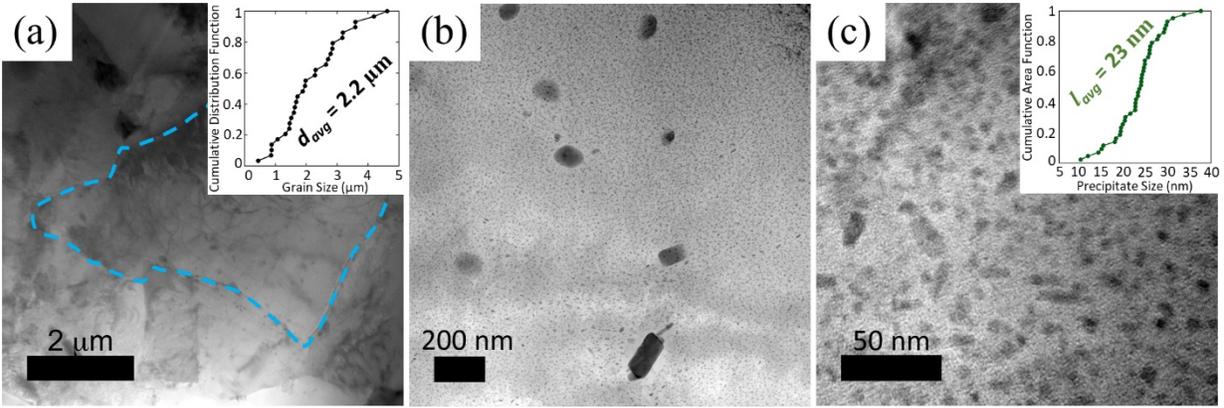

**Fig. 4** Bright field TEM micrographs of the Al matrix. (a) One Al grain is outlined in blue dashed lines, and the inset is a cumulative distribution function of the grain size with an average of 2.2 μm. (b) Precipitates of various sizes ranging from tens to hundreds of nanometers. (c) Nanoscale precipitates within the Al matrix material. The inset is a cumulative distribution function of the nanoscale precipitate size with an average of 23 nm

In addition to the intermetallics found by XRD, STEM-EDS also provided evidence of nanoscale MgO and Zn-rich phases that were primarily located near the matrix-particle interfaces (not shown here for brevity). The presence of $MgZn_2$ intermetallics was confirmed by STEM-EDS, as shown in Figure 5. These particles are tens of nanometers in diameter and distributed homogeneously throughout the Al matrix, in agreement with the data obtained from bright field imaging in Figure 4.

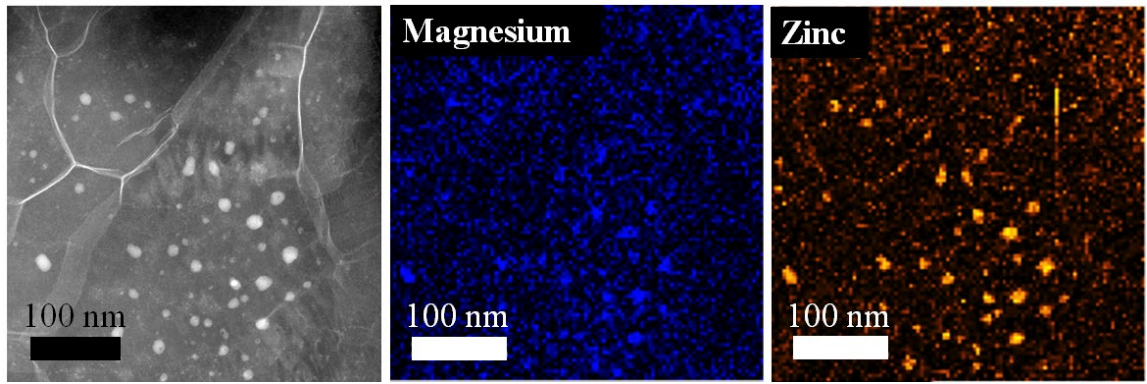

**Fig. 5** STEM-EDS maps showing the presence of nanoscale $MgZn_2$ intermetallics uniformly distributed within Al grains

Figure 6 shows the result of the microcompression test for Pillar A and the corresponding SEM snapshots at two strain values of 0.063 and 0.250. Pillar A has the most complex microstructure (Figure 1(a)), which contains one SiC particle at the top and another in the center, multiple Al grains below the top particle, and intermetallic precipitates of various sizes within the Al grains. The engineering stress-strain data are presented in Figure 6(a), where a yield strength of 430 MPa is obtained for this sample, consistent with reported values for SiC particle-reinforced Al-matrix composites [34,35]. After reaching the yield point, the stress-strain curve, if viewed by itself, suggests that the pillar might be strain softening. For Al-matrix composites, possible reasons for strain softening include strain localization and/or nucleation and growth of voids or microcracks [36]. However, no obvious shear band or voids/microcracks are observed in the SEM image corresponding to a strain of 0.063, as shown in Figure 6(c). Instead, Figure 6(d) shows that Pillar A undergoes bending and any apparent softening is due to this response, and is not an intrinsic feature of the plastic deformation of this material. We hypothesize that the bending is a result of the misalignment caused by elastic inhomogeneities, since the Young's modulus of the Al matrix (69 GPa, [37]) is dramatically lower than that of the SiC particles (410 GPa, [38]). No obvious separation of the complete particle from the matrix is observed from SEM images of the pillar after plastic deformation, although heavy deformation is found at some locations along the interface. Therefore, Pillar A with the most complex microstructure exhibits a yield strength of 430 MPa and can undergo stable plastic flow without fracture of the matrix-particle interface.

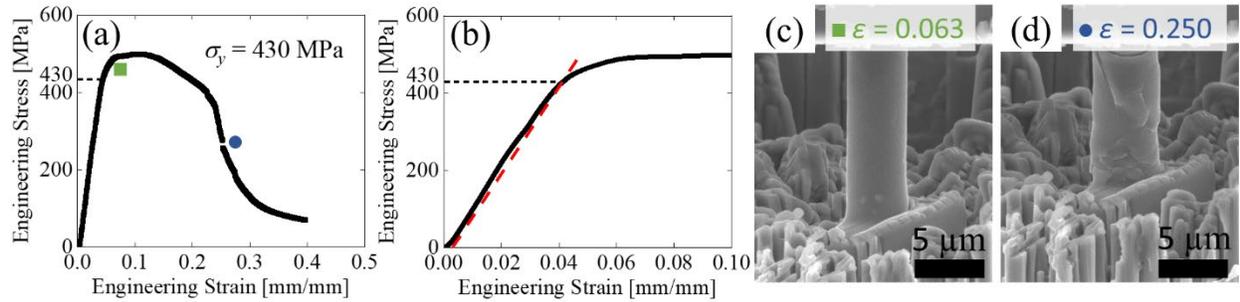

**Fig. 6 (a) Engineering stress-strain curve for Pillar A from microcompression testing, with a zoomed view provided in (b). The red dashed line represents the 0.2% offset with the yield strength value indicated on the Y axis in each as $\sigma_y$. (c) and (d) are SEM snapshots corresponding to engineering strains of 0.063 and 0.250, respectively**

The engineering stress-strain curve and two SEM snapshots for Pillar B are shown in Figure 7. Compared to Pillar A, Pillar B has a larger size but simpler microstructure with only one SiC particle at the top and one Al grain at the bottom region. Therefore, while Pillar A has multiple grain boundaries and the matrix was a polycrystal, no grain boundaries exist in Pillar B. In addition, the volume fraction of the SiC particle is higher in Pillar B than in Pillar A, and the matrix-particle interface terminates in the top surface of the pillar, as shown later in SEM and TEM images of the deformed pillar. Due to the different microstructure, the plastic deformation of Pillar B deviates from that of Pillar A. Figure 7(a) shows that Pillar B has a much lower yield strength of 290 MPa, after which a plateau occurs followed by apparent strain hardening. It is possible that the low yield strength is due to lack of grain boundary hardening. However, the microcompression experiment on Pillar C (reported below), which also does not contain any grain boundary yet exhibits the highest yield strength among the three pillars, rules out this possibility. At the early stage of plastic deformation, the pillar does not show any failure initiation (Figure 7(c)), such as microcracks or shear bands. When the applied strain becomes larger, the top right section of the pillar shears downwards, as clearly seen in Figure 7(d). Therefore, the apparent strain hardening after the plateau observed in Figure 7(a) is due to a quickly increasing contact area as the sheared section moves downward and new material is exposed, meaning both the original cross-sectional area plus some near region is being contacted (Figure 7(d)). The apparent hardening during the later deformation stages in Pillar B is thus geometric and not associated with any true material response.

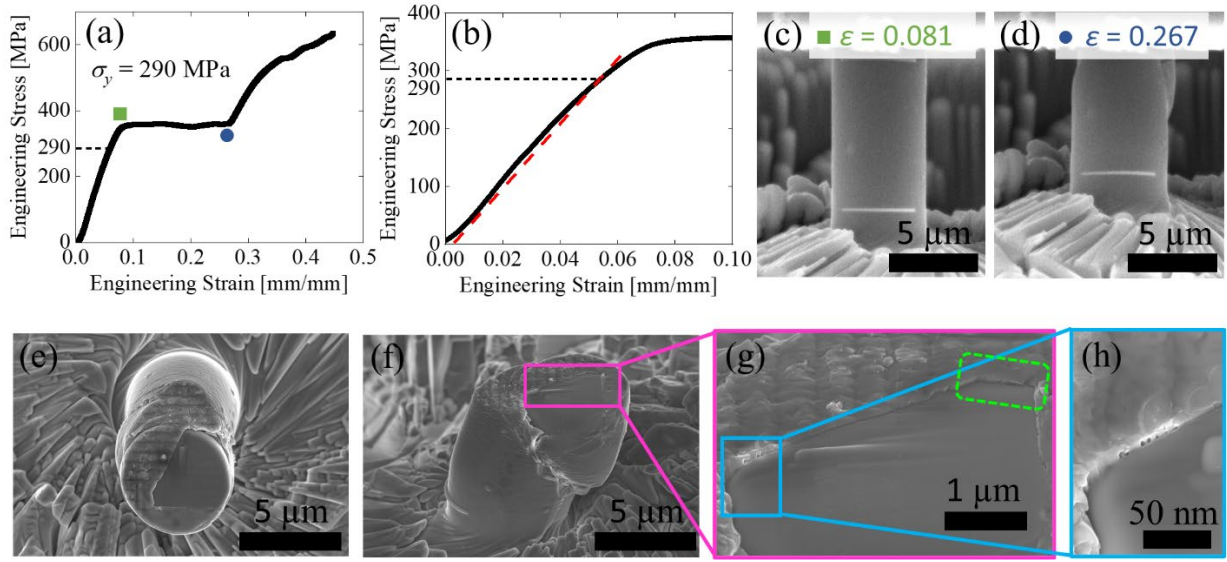

**Fig. 7 (a) Engineering stress-strain curve for Pillar B from microcompression testing, with a zoomed view provided in (b). The red dashed line represents the 0.2% offset with the yield strength value indicated on the Y axis in each as $\sigma_y$. (c) and (d) are SEM snapshots corresponding to engineering strains of 0.081 and 0.267, respectively. (e) Top view of the deformed pillar, which clearly shows matrix-particle interface debonding. (f) Side view of the deformed pillar demonstrating catastrophic failure through strain localization within a shear band. (g) A magnified view of the interface debonding area reveals two types of debonding. The first one is brittle with no/little plastic deformation in the Al matrix, as enclosed in the green dashed box. The second one is decohesion with plastic deformation in the matrix, as outlined in a blue rectangle and presented in greater magnification in (h)**

Figures 7(e) and 7(f) are the top and side views of deformed pillar B, respectively. Unlike Pillar A, which undergoes stable plastic deformation, Pillar B fails catastrophically through strain localization within a dominant shear band (Figure 7(f)). The shear band crosses from the top left to the middle right of the sample and is aligned closely with the matrix-particle interface, as shown more clearly in TEM images of the deformed pillar. The difference in deformation mode between Pillar A and B indicates that grain boundaries are effective obstacles for preventing shear localization. In addition to catastrophic failure, debonding at the matrix-particle is observed for the deformed pillar. To get a closer look at the interface debonding, a magnified view of the region outlined by a purple rectangular in Figure 7(f) is shown in Figure 7(g). At this top surface, two types of debonding are observed. The first type is a clean and brittle separation with no obvious deformation in the matrix, as outlined by a dashed green box. The second type is a decohesion with plastic flow in the matrix, which is enclosed by a blue rectangular in Figure 7(g), with a magnified view of this region shown in Figure 7(h). The occurrence of these two decohesion modes could be related to whether or not oxides and/or intermetallics are clustered near that particular interfacial region, but a detailed investigation of this effect is beyond the scope of this paper. Therefore, without multiple grain boundaries, Pillar B undergoes catastrophic failure through strain localization within a dominant shear band along with matrix-particle interface debonding. However, the relationship between the shear localization and interface debonding is not clear at this point.

Figure 8 shows the microcompression results for Pillar C, whose microstructure resembles a simple matrix-particle interface. Same as Pillar B, Pillar C contains one SiC particle and one Al grain. However, the volume fraction of SiC in Pillar C is much higher, being ~50 %, and the matrix-particle interface is across the whole sample from the top left to bottom right of the pillar (Figure 1(c)). Figure 8(a) displays the engineering stress-strain data for Pillar C, where a yield strength of 580 MPa is obtained. This is the highest yield strength among the three pillars, which indicates that grain boundary strengthening cannot be the primary strengthening mechanism for the Al matrix because Pillar C does not contain any grain boundaries. Consequently, other strengthening mechanisms should play a dominant role for the matrix and will be discussed below.

After the onset of yielding, Figure 8(c) shows that a faint shear band is formed at the bottom, but this strain localization is far away and orientated differently from the matrix-particle interface. With further increasing applied strain, Pillar C experiences catastrophic failure through a dominant shear band in the Al matrix, as shown in Figure 8(d). The dominant shear band is close to the initial shear band observed in Figure 8(c) and also far away from the matrix-particle interface. The reason that the dominant shear band forms in the Al matrix and far away from the interface may be higher dislocation density in the matrix than in the near interface region, as observed by Hong et al. [39] for a 1060 Al/Al$_2$O$_3$ composite after compression tests. This higher dislocation content can provide many easily moveable defects that can accumulate large plastic strains in a local region. Moreover, without any grain boundaries, there is an uninterrupted slip pathway that traverses the entire micropillar sample. The high yield strength of 580 MPa of Pillar C suggests that the Al matrix is intrinsically very strong.

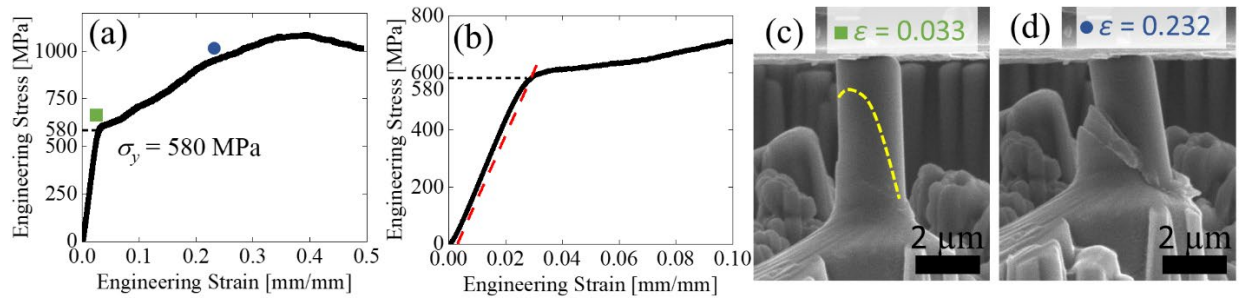

**Fig. 8 (a) Engineering stress-strain curve for Pillar C from microcompression testing, with a zoomed view provided in (b). The red dashed line represents the 0.2% offset with the yield strength value indicated on the Y axis in each as $\sigma_y$. (c) and (d) are SEM snapshots corresponding to engineering strains of 0.033 and 0.232, respectively. Only matrix plastic deformation is observed here**

It should be emphasized that even though both Pillar B and C exhibit catastrophic failure, the location and alignment of the dominant shear band with respect to the matrix-particle interface is not the same. The shear band in Pillar B is in the upper region of the pillar and aligned with the matrix-particle interface, while the one in Pillar C is at the bottom, far away from the interface, and at a different angle than the interface plane. Therefore, these differences between Pillar B and C may lead to different interface behavior after deformation. Unlike Pillar B, no interface debonding is observed for Pillar C. Hence, we propose that when the shear band occurs in the vicinity of the matrix-particle interface and aligns with it, as in the case of Pillar B, decohesion at the interface occurs. In contrast, if the shear band is far from the interface and does not align with any segment of the interface, the particle will not be separated from the matrix.

Since Pillars A and B exhibit the most contrasting plastic deformation (steady plastic flow without interface failure vs. catastrophic failure through shear localization with interface debonding), cross-sectional TEM samples were lifted out using FIB from both pillars after deformation to further investigate the relationship between microstructure and mechanical behavior. This information is shown in Figures 9 and 10. Figure 9(a) is a bright field TEM micrograph of the deformed Pillar A, where the top SiC particle and Al grains are outlined by yellow and blue dashed lines, respectively. At least four Al grains sit below the SiC particle, and the size of both the particle and grains is in the range of a few micrometers. Each grain has its own independent dislocation network and no strain localization occurs in this specimen, even at relatively high stress levels. Consequently, grain boundaries are shown to act as effective obstacles to prevent shear localization of the larger sample. As mentioned earlier, the shear localization in single-grain Pillars B and C may be due to a high density of mobile dislocations and an uninterrupted slip path through the sample, which leads to shear banding. For samples with multiple grains, dislocations need to transmit from one grain to another in order to form a shear band, so no one slip path can dominate. Shen et al. [40] calculated the stresses needed for dislocation transmission in a stainless steel, and the values obtained are higher than the macroscopic yield stress.

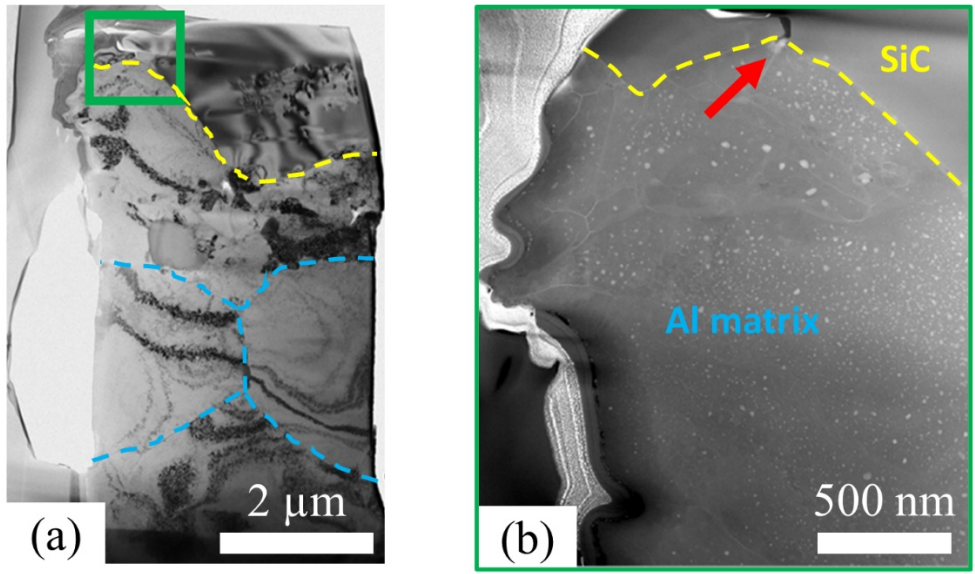

**Fig. 9 (a) Bright field TEM micrograph of the deformed Pillar A. SiC particle and Al grain boundaries are outlined by yellow and blue dashed lines, respectively. A region containing a crack that has arrested at the matrix-particle interface is enclosed by a green square and magnified view is shown in (b). The exact location of crack arrest is denoted by a red arrow at the matrix-particle interface**

A crack is observed that travels through the SiC particle and ends near the matrix-particle interface, as enclosed in a green rectangular in Figure 9(a) and magnified in Figure 9(b). Instead of propagating along the interface, the crack is arrested at the interface and does not lead to any obvious interface deformation. This implies that the matrix-particle interface is not an intrinsic weak point that is ready to fracture easily and to initiate macroscopic failure. By performing tensile tests on a SiC particle reinforced Al 6061 alloy after three different heat treatments with different cooling rates, Gupta et al. [41] concluded that one reason for an enhanced matrix-particle interface bonding can be a higher concentration of alloy elements at the interface than in the matrix. In the present study, we also observed some interface regions coated with alloy elements from EDS maps (not shown here for brevity), which may attribute to the strong interface observed in our composite material. In addition, the observation of crack arrests shows that crack propagation requires additional energy to move along the matrix-particle interface. Figure 9(b) also shows many nanosized precipitates within the Al matrix which are not heavily deformed after the microcompression test in this pillar. Therefore, for a bulk-like microstructure such as that captured by Pillar A, the grain boundaries can effectively prevent shear localization up to a relatively high stress, and failure does not originate from the matrix-particle interface.

When the microstructure of the pillar becomes simpler (e.g., only one Al grain below the SiC particle as seen in Pillar B and C), catastrophic failure through shear localization occurs. Moreover, Pillar B also exhibits matrix-particle interface debonding. Figure 10(a) shows a bright field TEM micrograph of deformed Pillar B, where most of the SiC particle has been removed during FIB milling to allow for an electron transparent sample and the rest of the particle is outlined by a dashed yellow line. No grain boundaries exist in the matrix, and several regions with strain contrast are observed in the bottom region. The top part of the pillar is sheared to the right with respect to the bottom region along a major shear band, which crosses the pillar from top left to middle right and is enclosed in a purple rectangular in Figure 10(a). Since no grain boundaries exist within the sample, once shear localization is initiated, the dominant shear band can propagate through the whole material without barriers and cause catastrophic failure. Figure 10(a) also clearly shows that the dominant shear band and part of the matrix-particle interface are aligned with each other, which is suspected to be related with interface debonding. A closer examination of the shear band in Figure 10(b) with dark field TEM imaging reveals a high dislocation density. In addition, a void was formed at the interface between a precipitate and matrix (denoted by a green arrow). Figure 10(c) presents a further magnification view of the shear band, which shows that the nanoscale precipitates are severely sheared in this region. We note that the nanoscale precipitates are not obviously deformed in Pillar A (Figure 9(b)), which exhibits a high yield strength, so we can conclude that the strain localization and the damaged precipitates lead to a lower yield strength for Pillar B.

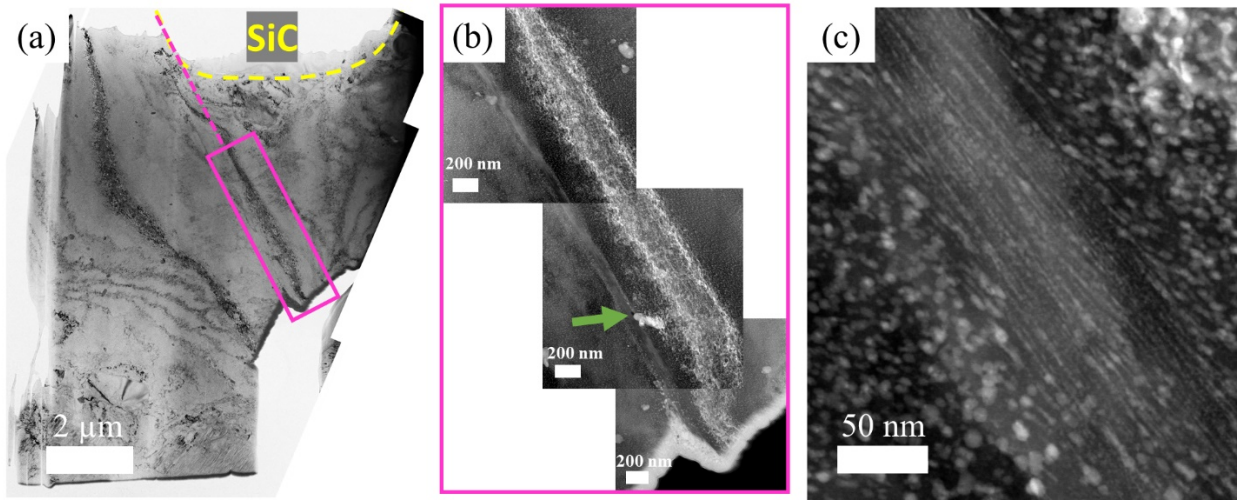

**Fig. 10 (a) Bright field TEM micrograph of the deformed Pillar B. The SiC particle is outlined by a dashed yellow line, which also represents the matrix-particle interface. The area outlined by a purple box shows a dominant shear band, which is shown in more detail in (b). The dotted purple line follows the trace of the dominant shear band, which aligns well with the matrix-particle interface. (b) The dislocation density within the shear band is very high, and a void has formed at the interface between an intermetallic precipitate and matrix, as denoted by a green arrow. (c) A magnified view of the shear band demonstrates that the precipitates are severely deformed in this region**

The microstructure evolution during plastic deformation for the three pillars indicates that strain localization and matrix-particle interface decohesion are related. However, whether interface debonding precedes and causes shear localization or shear localization facilitates interface debonding still needs to be determined. To clarify the relationship, in situ direct debonding testing was next employed, where load is applied on one SiC particle to separate it from the matrix and SEM observation allows the exact moment of debonding to be isolated. Before applying the load, two clean surfaces perpendicular to each other with SiC particles on both the top and front surfaces are created using FIB milling, as shown in Figure 11(a). The dark and light regions are SiC particles and Al matrix, respectively. Next, a slightly rounded tip that is smaller than the particle is used to apply load on one particle until decohesion at the particle-matrix interface occurs. The primary output from the direct debonding experiments is the critical applied load when debonding occurs. Figure 11(b) displays two SEM snapshots, with the first being when the tip has just contacted the particle and the second being when the particle was pushed out from the matrix. The corresponding displacements of the tip are indicated. The SEM images clearly show that only one SiC particle is separated from the matrix, so the interface strength can be obtained via the applied load corresponding to interface debonding and the measured particle size. Figure 11(c) is a representative force vs. displacement curve, where a drop in the force occurs at a displacement of ~1.7 μm. Without in situ imaging, it would be likely that one would associate this drop with the initiation of the debonding event. However, SEM snapshots show that the debonding does not occur until a larger displacement of ~2.2 μm with an applied force of ~2900 μN. This point is marked by a red circle in the force-displacement data, and an SEM micrograph after particle debonding is displayed to the right. No obvious drop in the load is observed when the particle debonds from the matrix. This demonstrates that in situ direct debonding testing is a powerful technique to allow for a precise determination of the load associated with the interface decohesion.

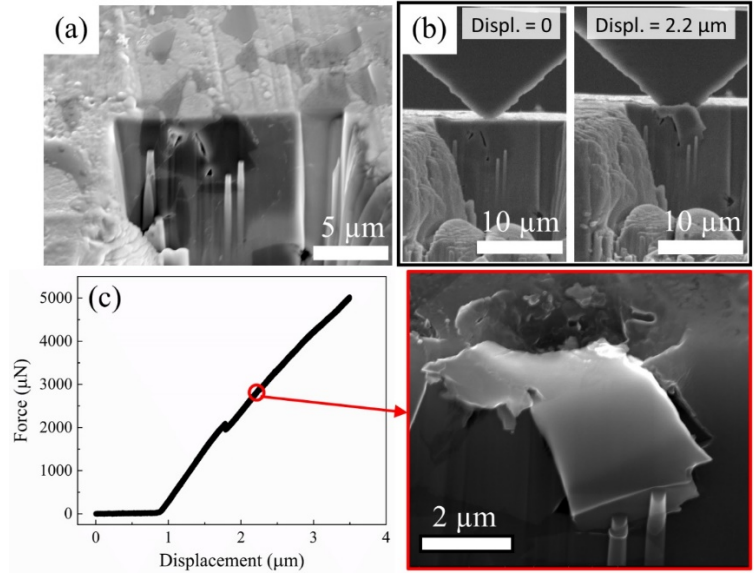

**Fig. 11 (a) Two perpendicular surfaces containing one SiC particle are created using FIB milling. (b) A load is applied on the particle using a slightly rounded tip until the particle separates from the matrix. The corresponding displacements are shown for the two images. (c) Representative force-displacement data from a direct debonding test. The obvious drop in the force does not correspond to the initiation of the interface debonding, which actually takes place at a later displacement marked by the red circle. The corresponding SEM micrograph of the debonded particle shows no obvious plastic deformation of the matrix near the interface**

After determining the load associated with the interface debonding, particle sizes must be measured in order to estimate the interfacial area and calculate the interface strength. It is important to recall the goal of this experiment, which is the identification of the weak point within the microstructure that results in failure. Since we know the resolved shear stress for failure in Pillar B, where both strain localization and particle debonding occur, we want to compare this with the measured strength of simple interface debonding. Due to the irregular shape of the SiC particles tested and lack of information about the internal debonding process inside the material, we choose to use a time-intensive method for determining the interfacial area of the particles and a conservative procedure for calculating interface strength. As a whole, we perform measurements and make assumptions that mean we are certainly overestimating the matrix-particle interfacial area during the debonding experiments, which translates into the determination of a lower measurement bound for the interface strength. We therefore know that the real interface strength is at least as high as (and likely higher than) the value calculated here.

First, the shape of the particle observed on the surface during the experiment is not necessarily representative of the internal particle shape. For example, the particle cross-section can be larger or smaller underneath the surface than on the surface. To address this potential issue, particle sizes are determined by post-test milling using FIB to different depths, with SEM images of the shapes that particles left behind taken at each depth, as illustrated in Figure 12. For the SEM image with no milling depth, $z_0 = 0$, a rectangle is used to enclose the particle shape on the top surface, and the side lengths are $L_1^{(0)}$ and $L_2^{(0)}$. Next, the top surface is milled down by a depth of $z_1$ nm, and another rectangle with side lengths of $L_1^{(1)}$ and $L_2^{(1)}$ is employed to measure the particle size. After $n$ milling steps, the side lengths of the rectangular used for measurement are $L_1^{(n)}$ and $L_2^{(n)}$. For each depth, a corresponding effective side length is defined as $L_{eff}^{(i)} = (L_1^i \times L_2^i)^{1/2}$. To be conservative, the particle shape is assumed to be a cube with a side length, $L_{eff}^{avg}$, equal to $(1/n) \times \sum_i L_{eff}^{(i)}$, $i = 1, \ldots, n$. Finally, the total interface area between the particle and matrix is set to be $4 \times (L_{eff}^{avg})^2$. The assumption of four sides in contact here acknowledges that the front and top of the particle are free surfaces created by FIB milling. A second potential issue that can arise is that one matrix-particle interface can debond before the others, due to variations in local stresses, particle shape, or even interfacial strength. The effect of this would mean that the actual area of the interface that debonds is less than the total interfacial area of the particle, and the true interface strength higher than one that is calculated using that total interface area. However, to be conservative, we assume that all four faces of the particle debond at the same time, which provides a lower bound on the interface strength.

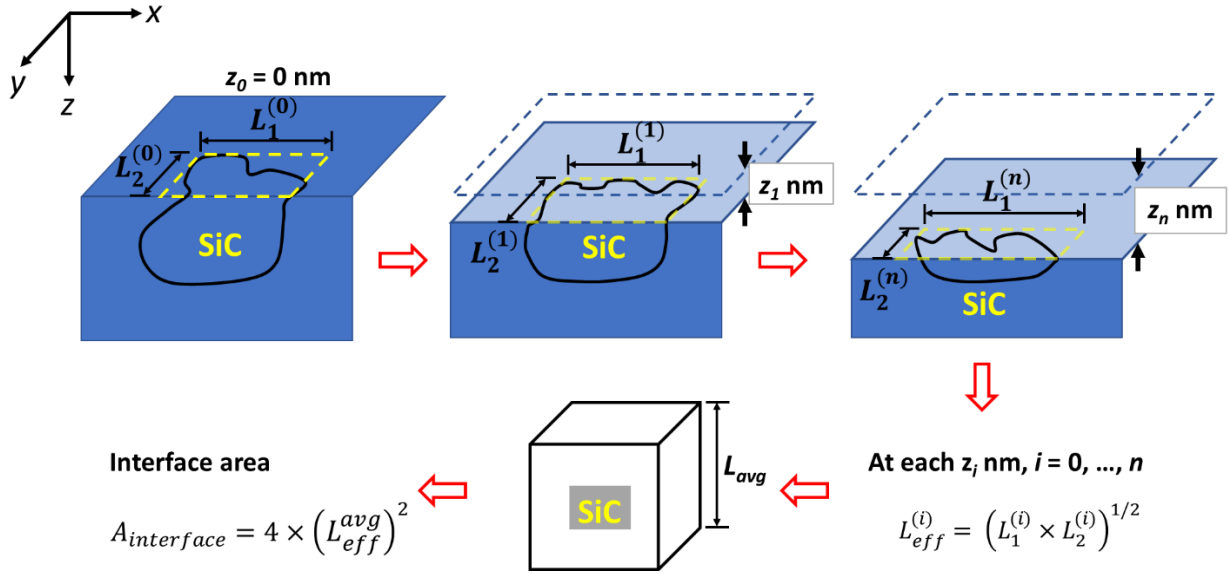

**Fig. 12 Schematic illustration of measuring particle sizes via post-test milling at different depths. A rectangular is used to enclose the shape of the particle on the top surface (x-y plane) at each depth $z_i$, and the side lengths are $L_1^{(i)}$ and $L_2^{(i)}$. Then, an effective length at each $z_i$, $L_{eff}^{(i)}$, is obtained by taking the square root of the product of $L_1^{(i)}$ and $L_2^{(i)}$. The shape of the particle is assumed to be a cube with a side length of $L_{eff}^{avg}$, equal to the average value of $L_{eff}^{(i)}$ at all depths. Four sides of the particle are assumed to be contacted with the matrix. Therefore, the total interface area, $A_{interface}$, is four times the square of the $L_{eff}^{avg}$**

Figure 13 shows representative measurements for two particles, with (a)-(f) corresponding to a milling depth of 50 -300 nm. All SEM snapshots clearly show two particles, indexed as "1" and "2" and outlined by dashed rectangles. In addition, measurements of the side lengths, $L_1$ and $L_2$, are shown for each particle. It should be noted that the SEM images have a tilt angle of 52 degrees, and corrections have been made when measuring the $L_2$ length values. Next, the $L_{eff}$ value for each particle and milling depth is obtained by taking the square root of the product of $L_1$ and $L_2$. Subsequently, an average is taken of $L_{eff}$ values for all milling depths, and this average value is used to calculated the effective area, as described above, and calculate the interface strength of a specific particle. Six particles were tested and measured in this way, and their results are listed in Table 1, giving an average interface strength of 254 ± 61 MPa. This value is much larger than the critical resolved shear stress for Pillar B of ~95 MPa, calculated using an average Schmid factor of 0.3 for face-centered cubic Al [42]. In fact, all individual measurements are much larger than this critical resolved shear stress. Again, we remind the reader that our calculation of the interface strength is a lower bound, and that the true interface strength is at least this value and likely higher. As a whole, these results indicate that the matrix-particle interface is not a weak point to start failure and initiate strain localization. In fact, our results show that shear localization should be occurring first and can facilitate interface debonding if intense shear banding occurs.

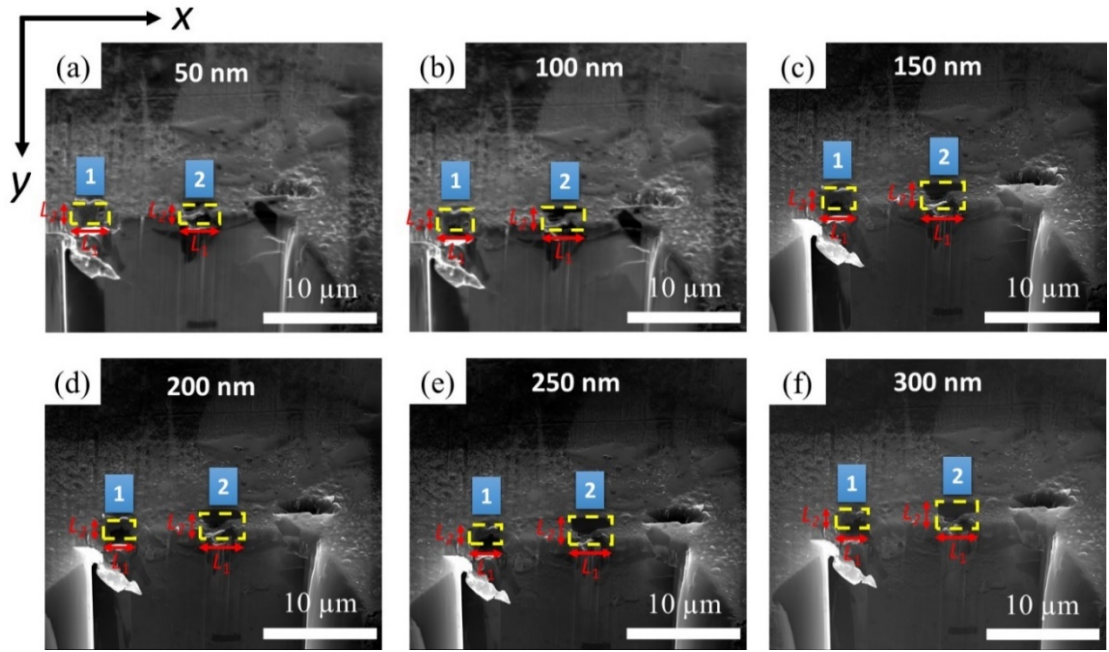

**Fig. 13 Representative particle size measurements for two particles, labeled as "1" and "2", at different milling depths, as indicated in each SEM image. Dashed yellow rectangles are used to enclose the shapes that the debonded particles left behind. $L_1$ and $L_2$ correspond to the horizontal and vertical side lengths of the particles, respectively**

**Table 1.** $L_1$, $L_2$, and $L_{effective}$ for each tested particle as well as the corresponding interface strength for each particle. The average interface strength is 254 ± 61 MPa.

| Particle Index | $L_1^{avg}$ [μm] | $L_2^{avg}$ [μm] | $L_{eff}^{avg}$ [μm] | Interface Strength [MPa] |
|---|---|---|---|---|
| 1 | 2.46 | 1.80 | 2.10 | 207 |
| 2 | 3.24 | 2.44 | 2.81 | 189 |
| 3 | 2.14 | 2.37 | 2.24 | 320 |
| 4 | 1.98 | 2.30 | 2.13 | 303 |
| 5 | 3.44 | 2.70 | 3.03 | 200 |
| 6 | 2.70 | 2.22 | 2.44 | 308 |
| Avg. | | | | 254 |
| Std. Dev. | | | | 61 |

Due to the importance of matrix-particle interfaces on the performance of metal-matrix composites, various approaches have been employed to study the interfacial characteristics. For example, Guo *et al.* [43] deliberately fabricated micropillars with a single SiC/Al interface at 45° angle with respect to the loading axis, and then performed uniaxial compression tests. They obtained an interfacial shear strength of 133 ± 26 MPa, which is not far from the value in the present study considering the pure Al in their study vs. Al alloy in the present study, and both values fall into a range predicted by numerical simulations based on cohesive zone model [44,45]. Hahnlen and Dapino [46] studied the shear strength of fiber-matrix interfaces in an Al 3003-H18 matrix with embedded prestrained NiTi fibers that was fabricated through ultrasonic additive manufacturing. By combining differential scanning calorimetry technique and constitutive modeling for thermally-induced strain of composites, these authors obtained an average interface shear strength of 7.28 MPa. This value is much lower than the interface strength obtained in the present study, which might be partly due to the dramatically different length scales of fibers (a few to tens of mm) in their study versus SiC particles (a few μm) in the present study. The wettability between different types of materials (Al and NiTi in Ref. [46] versus Al and SiC in the present study) could be another possible reason for the significantly different interface strengths. Later, Hehr and Dapino [47] employed both fiber pullout tests and finite element analysis to study the matrix-fiber interfacial shear stress and failure behavior for the same material. These authors identified the Al matrix to be the weakest link in the composite in their study. The reason for matrix failure instead of interface decohesion is that a large area of recrystallized Al grains form near the interface during the fabrication process and the sizes of these recrystallized grains are submicron and micron, which are similar in length scale to the fiber surface asperities. As a result, submicron grains were trapped within the asperities of the fiber, leading to a robust mechanical interlocking and high resistance to shear at the matrix-fiber interface. In the present study, the matrix-particle interfaces also turn out to be strong. In addition, a more thorough characterization of the relationship between matrix failure and interface debonding is provided, which depends on the alignment and position between the dominant shear localization and interface.

The present study not only offers a detailed characterization of the plasticity and failure of an Al matrix composite on multiple length scales, such as severely deformed nanoscale intermetallic precipitates to shear localization within micrometer-sized grains, but also sheds light on paths to designing improved composite materials. For instance, our experiments show that grain boundaries are effective obstacles for preventing shear localization, and that this localization must be avoided because it can cause matrix-particle debonding and subsequent failure at low stress. Hence, decreasing the grain size (and therefore having more grain boundaries) should be beneficial for restricting strain localization. There is likely an interplay between grain size and reinforcement particle size too with such a mechanism, as grains smaller than the added reinforcement particles should be preferred to ensure that there are multiple grain boundaries near each matrix-particle interface. Indirect evidence for such materials design can be found in the literature, where high-performance Al matrix composites with nanosized grains smaller than 45 nm have been successfully fabricated using high-energy ball milling process [48]. In addition to reducing the crystallite size, many studies focus on decreasing the size of reinforcements for improving the mechanical strength of the composite. For example, Prabhu et al. [49] synthesized an $Al_2O_3$ reinforced Al matrix composite with a particle size of 50 nm. However, the present study shows that caution should be exercised when using smaller particles, since although this size reduction might increase strength, it may also result in early interface debonding if a dominant shear band is formed. Consequently, a trade-off needs to be considered when choosing the optimal size of reinforcement. Recently, novel manufacturing techniques have been employed to enhance the mechanical properties of Al matrix composites. For instance, by injecting milled particulate Al-SiC$_p$ composite powders into Al356 melt and then compocasting into bulk samples, Amirkhanlou and Niroumand [50] found that the uniformity of the SiC particles was significantly improved with both the grain size and particle size decreased, which results in an increase in hardness and bending strength and strain. Finally, the intermetallic precipitate size is another factor to be considered since these features were found to control the strength of the Al matrix in this study. Therefore, tuning the absolute and relative sizes of matrix grains, reinforcement particles, and intermetallic precipitates should allow a balance to be reached in the design of improved composite materials.

## 4. Conclusions

The present study offers valuable insights of the deformation behavior and failure modes of SiC particle-reinforced Al matrix composite on multiple length scales, by employing both in situ small-scale mechanical testing and advanced electron microscopy. Catastrophic failure through shear localization can be effectively prevented by Al grain boundaries, and the dominant strengthening effect comes from intermetallic precipitates within the Al grains. When shear localization occurs, it can drive subsequent matrix-particle interfacial debonding if the shear band is close to and aligns with part of the interface. If the shear band is far away from the interface and does not align with the interface, the SiC particle will not separate from the Al matrix. More importantly, the matrix-particle interface is not intrinsically a weak point for failure initiation.

## Data availability

The raw/processed data required to reproduce these findings cannot be shared at this time as the data also forms part of an ongoing study.

## Acknowledgements

The authors would like to acknowledge financial support from the BIAM-UCI Research Centre for the Fundamental Study of Novel Structural Materials (Research Agreement #210263).

## Competing interests

The authors declare that they have no competing interests that could have appeared to influence the work reported in this paper.